\documentclass[prd,preprint,a4paper,superscriptaddress,nofootinbib,11pt]{revtex4}
\usepackage{amsmath}
\usepackage{amssymb}
\usepackage{amsthm}
\usepackage{stmaryrd}
\usepackage{url}
\usepackage{mathrsfs}
\usepackage{textcomp}
\usepackage{amsfonts}
\usepackage{txfonts}
\usepackage{eufrak}
\usepackage[sans]{dsfont}

\def\fs{\footnotesize}
\begin{document}

\title{\Large $\kappa$-Minkowski spacetime as the result of Jordanian twist
deformation}
\author{A. Borowiec\footnote{borow@ift.uni.wroc.pl, borovec@theor.jinr.ru}}

\affiliation
{\fs
Institute for Theoretical Physics, University of Wroclaw,
pl. Maxa Borna 9, 50-205 Wroclaw, Poland\\}
\affiliation
{\fs Bogoliubov Laboratory of Theoretical Physics,
Joint Institute for Nuclear Research, Dubna,
Moscow region 141980, Russia}
\author{A. Pachol\footnote{anna.pachol@ift.uni.wroc.pl}}
\affiliation{{\fs
Institute for Theoretical Physics, University of Wroclaw,
pl. Maxa Borna 9, 50-205 Wroclaw, Poland}}


\begin{abstract}
Two one-parameter families of twists providing $\kappa-$Minkowski $*$ -product deformed spacetime are considered: Abelian and Jordanian.
We compare the derivation of quantum Minkowski space from two perspectives.
The first one is the Hopf module algebra point of view, which is strictly
related with Drinfeld's twisting tensor technique. The other one relies on an
appropriate extension of "deformed realizations" of nondeformed Lorentz
algebra by the quantum Minkowski algebra. This extension turns out to be de
Sitter Lie algebra. We show the way both approaches are related. The second
path allows us to calculate deformed dispersion relations for toy models
ensuing from different twist parameters. In the Abelian case one
recovers $\kappa-$Poincar\'{e} dispersion relations having numerous applications
in doubly special relativity. Jordanian twists provide a new type of dispersion relations which in
the minimal case (related to Weyl-Poincar\'{e} algebra) takes an
energy-dependent linear mass deformation form.
\end{abstract}

\maketitle
\section{Introduction}

Noncommutative geometry has found many applications in physical theories in
recent years. It has been suggested as a description of spacetime at the Planck
scale and proposed as a background for the unification of gravity and quantum
field theory. The first idea of non commuting coordinates
was suggested as long ago as the 1940s by Snyder \cite{Snyder}. More recently,
deformed coordinate spaces based on algebraic relations $[x^\mu,x^\nu]=i%
\theta^{\mu\nu}$ (with $\theta^{\mu\nu}$ constant) have been introduced in Ref.
\cite{Dop1} as a consequence to quantum gravity \cite{Dop1}-\cite{Dop2}.
These presently known as "canonical" spacetime commutation
relations have been the subject of many investigations (see, e.g., \cite{Dop1}-%
\cite{ABDMSW}).
Alternatively, this type of noncommutative coordinates was introduced in
string theory as coordinates on the spacetime manifold attached to the ends
of an open string in a particular gauge field background \cite{SW}. The same
authors have also introduced the so-called Seiberg-Witten map relating gauge
theories on commutative and noncommutative (NC) spaces. This research was a
beginning of extensive studies on quantum field theories (QFT) defined
over NC spaces \cite{DN}-\cite{Zupnik}, particularly with the twisted
($\theta-$deformed) Poincar\'{e} invariance of NC QFT in Refs. \cite{Chai1,Chai2}.
A growing number of investigations concern gauge theories \cite{MSSW}-\cite{ADMSW}
and, moreover, NC gravity \cite{ADMW}-\cite{Chaichian}.

The Lie-algebraic type of noncommutativity has also been widely investigated.
Inspired by the $\kappa$-deformed Poincar\'{e} algebra \cite{Luk1,Luk2},
the $\kappa$-Minkowski spacetime has been introduced in Refs. \cite{Z,MR},
together with dimensionfull masslike deformation parameter $\kappa$
(usually connected with Planck mass $M_P$). It has been furt her used by many
authors \cite{Luk1}-\cite{BACKG} as a starting point to construct quantum
field theories and then to discuss Planck scale physics. The past few years have
experienced an ever-growing interest in $\kappa$-deformed spacetime
motivated by so-called doubly special relativity (DSR) \cite{AC,BACKG} which includes,
besides the velocity of light, a second invariant mass
parameter ($\kappa$). In the framework of DSR, the kinematic consequences of
deformed spacetime have been examined; see, e.g., \cite{Smolin}-\cite{Gosh}.

Lie-algebraic quantum deformations (\ref{c1}) are the most physically appealing from a
bigger set of quantum deformations (\ref{tcr}) in the Hopf-algebraic framework of
quantum groups  \cite{Drinfeld}-\cite{Fadeev}. It appears that quantum deformations of Lie
algebras are controlled by classical $r$ -matrices satisfying the classical
Yang-Baxter (YB) equation: homogeneous or inhomogeneous.
Particularly, an effective tool is provided by the so-called
twisted deformations \cite{Drinfeld} which classical $r$ matrices satisfy the
homogeneous YB equation and can be applied to both Hopf algebra
(coproduct) as well as related Hopf module algebra ($*$ product) \cite{Oeckl}.
Two types of explicit examples of twisting two tensors are the best known and investigated in
the literature: Abelian \cite{Reshetikhin} and Jordanian \cite{Ogievetsky} as
well as the extended Jordanian one \cite{Kulish} (see also \cite{Bonneau,VNT1}). The $\kappa$ -deformation
of Poincar\'{e} algebra is characterized by the inhomogeneous classical YB equation, which
implies that one should not expect to get $\kappa-$Minkowski space from a
Poincar\'{e} twist. However, twists belonging to extensions of the Poincar\'{e} algebra are not
excluded \cite{Bu,MSSKG}. An interesting result has been found in Ref.
\cite{Ballesteros}, 
where starting from nonstandard (Jordanian) deformed $D=4$ conformal
algebra the $\kappa-$Minkowski space has been obtained in two ways: by
applying the Fadeev-Reshetikin-Takhtajan technique \cite{Fadeev} or exploiting a bialgebra structure
generated by a classical $\mathfrak{so}(4,2)$ $r$ matrix.

As a final remark, let us remind the reader that the classification of quantum
deformations strongly relies on classification of the classical $r$ %
matrices. In a case of relativistic Lorentz and Poincar\'{e} symmetries, such a
classification 
has been performed some time ago in Ref. \cite{Zakrzewski}. A passage from the
classical $r$ -matrix to twisting two-tensor and the corresponding Hopf-algebraic deformation is a nontrivial task. Explicit twists for
Zakrzewski's list have been provided in Ref. \cite{Varna} as well as their
superization \cite{VNT1,VNT2}. 
Systematic quantizations of the corresponding Lorentz and
Poincar\'{e} Hopf algebras are carried on in Ref. \cite{BLT2}. In particular, the
noncommutative spacetimes described by three types of Abelian Poincar\'{e}
twists have been calculated in Ref. \cite{LW}. 

In this article we shall work with a flat spacetime $\mathds{M}\equiv %
\mathds{R}^n$ of arbitrary dimension $n\geq2$. The paper is organized as
follows: in Sec. II, we establish formalism and notation.
We also show how to associate to a given $*$ -product the corresponding left
and right operator realizations of noncommutative coordinates by means of
generalized differential operators. Section III describes two families of
twists providing $\kappa$-Minkowski deformed spacetime with deformed
coproducts and antipodes. These are Abelian and Jordanian families. The Abelian
family has been previously investigated \cite{Bu}. One finds out, in an
explicit form, the operator realization of the $\kappa$-Minkowski coordinates $%
\hat x^\mu$ for each value of twist parameters. In this section we also
point out the smallest possible subalgebras, containing Poincar\'{e} algebra, to
which one can reduce the deformation procedure.
Lie algebra comprising the $\kappa$-deformed Minkowski algebra and deformed
realizations of the standard Lorentz algebra is introduced in Sec. IV.
This turns out to be $\mathfrak{so}(n, 1)$ algebra.
Similarly as in Ref. \cite{MS} one introduces deformed generators $M_{\mu\nu}$,
d'Alambert operator $\tilde\square$, and Dirac derivatives $D_\mu$ in the Lorentzian case.
It should be noted that in Ref. \cite{MS} it was Euclidean algebra. The generalized d'Alambert
operator allows one to calculate, in explicit form, dispersion relations for
plane wave solutions. A number of toy models based on different values of
twist parameters are considered. Appendix A collects improved ordinary
differential equations (ODE) for functions describing deformed generators.
Appendix B refers to Weyl-Poincar\'{e} algebra as to minimal algebra containing
both the Poincar\'{e} subalgebra as well as the twist element.

\section{Preliminaries and notation}

It is well -known that representations of Lie algebras (Lie groups) bring
these objects into the broader context of operator algebras which are essential
for quantum theories. In a case when representation space is, like in any
field theory, a space of functions (fields), this context is provided by an
algebra of (partial) differential operators on an underlying spacetime
manifold. In this case the algebra of functions itself becomes automatically
a Hopf module algebra over Hopf algebra of differential operators. Twisted
deformations in this broader context are a step forward in the geometrization of the
traditional Drinfeld scheme \cite{Drinfeld}.
It has been argued in a seminal paper \cite{ADMW} (see also \cite{ABDMSW,Oeckl,Aschieri}) that such a framework is very
useful in noncommutative geometry and deformed field theories and gives a hope
to describe gravity at the Planck scale \footnote{See, e.g., \cite{Chaichian} for a different framework.}.

As an example, let us consider Lie algebra $\mathfrak{igl}(n,\mathds{R})=%
\mathfrak{gl}(n,\mathds{R})\niplus\mathfrak{t}^{n}$ of the inhomogeneous
general linear group as a semidirect product of $\mathfrak{gl}(n,\mathds{R})$
with translations $\mathfrak{t}^{n}$. We choose a basis ${L_{\nu }^{\mu
},P_{\mu }}$ in $\mathfrak{gl}(n,\mathds{R})\niplus\mathfrak{t}^{n}$ with
the following standard set of commutation relations:
\begin{equation}  \label{igl}
\ [L_{\nu }^{\mu },L_{\lambda }^{\rho }]= \delta _{\nu }^{\rho }L_{\lambda
}^{\mu }-\delta _{\lambda }^{\mu }L_{\nu }^{\rho }~,\qquad\quad\; \ [L_{\nu
}^{\mu },P_{\lambda }]=-\delta _{\lambda }^{\mu }P_{\nu }
\end{equation}%
$\mu ,\nu , \ldots=0,\ldots,n-1$,\ and $n$ -denotes a dimension of spacetime $%
\mathds{M}=\mathds{R}^{n}$ which is not yet provided with any metric
structure. Nevertheless, for the sake of future applications, we shall use
"relativistic" notation with spacetime indices $\mu$ and $\nu $ running $%
0,\ldots,n-1$ and space indices $j,k=1,\ldots,n-1$.

This Lie algebra contains several interesting subalgebras, e.g., $\mathfrak{%
isl}(n,\mathds{R})=\mathfrak{sl}(n,\mathds{R})\niplus\mathfrak{t}^{n}$ of
the inhomogeneous special linear transformations. In this case instead of
diagonal generators $L_{\mu }^{\mu }$ we shall use its traceless
counterparts:
\begin{equation}  \label{central}
L_{\mu }= L_{\mu }^{\mu }- \frac{1}{n}L ,\qquad \mbox{where} \qquad L=\sum
L_{\mu }^{\mu }
\end{equation}
denotes a central element in $\mathfrak{gl}(n,\mathds{R})$,\footnote{%
However, it is not central in $\mathfrak{igl}(n,\mathds{R})$.} with $\sum
L_{\mu }=0$. Therefore, we have a basis $\{\ L_{\nu }^{\mu }, \mu \neq \nu ,
L_{k}, k=1,..,n-1, P_{\lambda }\ \}$ of $n^{2}+n-1$ elements in $\mathfrak{%
isl}(n,\mathds{R})$.

For any (constant of arbitrary signature) metric tensor
$g=(g_{\mu \nu })$: $g_{\mu \nu }=g_{\nu \mu }$
on $\mathds{R}^{n}$, one can associate a subalgebra of the
inhomogeneous orthogonal transformations $\mathfrak{iso}(g; n)=\mathfrak{so}%
(g; n)\niplus\mathfrak{t}^{n}\subset \mathfrak{isl}(n,\mathds{R})$, \footnote{%
We shall write $\mathfrak{iso}(n-p, p)$ whenever the signature $p$ will become
important.} which is defined by the following set of commutation relations:
\begin{equation}  \label{isog1}
\ [M_{\mu \nu },M_{\rho \lambda }]=g_{\nu \rho }M_{\mu \lambda }+g_{\mu
\lambda }M_{\nu \rho }-g_{\nu \lambda }M_{\mu \rho }-g_{\mu \rho }M_{\nu
\lambda }
\end{equation}%
\begin{equation}  \label{isog2}
\ [M_{\mu \nu },P_{\lambda }]=g_{\nu \lambda }P_{\mu }-g_{\mu \lambda
}P_{\nu }
\end{equation}
where $M_{\mu \nu }=-M_{\nu \mu }$ is defined by the embedding
\begin{equation}  \label{isog3}
\ M_{\mu \nu }=g_{\mu \lambda }L_{\nu }^{\lambda }-g_{\nu \lambda }L_{\mu
}^{\lambda }.
\end{equation}
The algebra $\mathfrak{igl}(n,\mathds{R})$, as well as its classical
subalgebras, acts on the algebra $\mathcal{A}_{\mathds{M}}$ of smooth
(complex-valued) functions on the spacetime manifold $\mathds{M}=\mathbb{R}%
^{n}$ via first-order differential operators (derivations$\equiv$vectors
fields): this is defined by natural representation - the so-called Schwinger
realization:
\begin{equation}  \label{schwinger}
\ L_{\nu }^{\mu }=x^{\mu }\partial _{\nu } \ \ , \ \ \ P_{\lambda}=\partial
_{\lambda}
\end{equation}
of $\mathfrak{igl}(n,\mathds{R})$ into infinite dimensional Lie algebra of
complex-valued vector fields $\mathcal{X}\mathds{M}=Der(\mathcal{A}_{\mathds{M}})$ . For the
purpose of deformations, one needs to work over a (algebraically closed) field
of complex numbers $\mathds{C}$. In what follows, we shall use complex Lie
algebras $\mathfrak{igl}(n)\equiv\mathfrak{igl}(n,\mathds{C})$, etc.,
instead their real counterparts. A real Lie algebra structure can be
eventually encoded in a corresponding reality structure (involutive
antiautomorphism) but we shall not focus on this point here.

In order to simplify the notation, we shall use the same letter to denote an
abstract element in $\mathfrak{igl}(n)$ and its (first-order) differential
operator realization in $\mathcal{X}\mathds{M}$ . This induces an embedding of the
corresponding enveloping algebras
\begin{equation}  \label{embed}
\ U_{\mathfrak{igl}(n)}\hookrightarrow U_{\mathcal{X}\mathds{M}}
\end{equation}
as an embedding of Hopf algebras with primitive coproducts
\begin{equation}
\Delta(X)=X\otimes 1+1\otimes X
\end{equation}
for $X\in \mathcal{X}\mathds{M}$. Nondeformed counit and antipode maps read as $%
\varepsilon(X)=0,\ \varepsilon(1)=1,\ S(X)=-X$, and $S(1)=1$. Let us emphasize that the enveloping algebra $%
U_{\mathcal{X}\mathds{M}}$ is simultaneously an algebra of linear (complex-valued,
partial) differential operators over $\mathds{M}$. It can be equipped with a
natural Hermitian involution defined on generators by $(x^\mu)^*=x^\mu$, $%
(\partial_\mu)^*=-\partial_\mu$. 
The embedding (\ref{embed}) provides a real (anti-Hermitian) realization ($%
Y^*=-Y$, for $Y\in \mathfrak{g}$ ) only for the subalgebra $\mathfrak{isl}(n,%
\mathds{R})$ and its subalgebras like $\mathfrak{iso}(g; n)$. In these cases
Hermitian conjugation is compatible with the corresponding reality
structures. The action via derivations of $\mathcal{X}\mathds{M}$ on $\mathcal{A}_{\mathds{M}}$
extends to a Hopf module algebra action of $U_{\mathcal{X}\mathds{M}}$ on the algebra
$\mathcal{A}_{\mathds{M}}$ (for details concerning Hopf module algebras, see, e.g., \cite{ABDMSW,ADMW}).

Since $\mathcal{A}_{\mathds{M}}$ is a Hopf module algebra over $U_{\mathcal{X}\mathds{M}}$, it
becomes automatically a Hopf module algebra over its sub-Hopf algebras, too,
particularly over $U_{\mathfrak{igl}(n)}$ as well as its subsequent Hopf subalgebras.
This can be further deformed, by a suitable twisting element $\mathcal{F}$,
to achieve deformed Hopf module algebra $(\mathcal{A}_{\mathds{M}}^{\mathcal{F}},U_{%
\mathcal{X}\mathds{M}}^{\mathcal{F}})$, where the algebra $\mathcal{A}_{\mathds{M}}^{\mathcal{F%
}}$ is equipped with a twisted star -product
\begin{equation}  \label{tsp}
f\star g=\mu\circ\mathcal{F}^{-1}(f\otimes g)=\bar{\mathrm{f}}^\alpha(f)\bar{%
\mathrm{f}}_\alpha(g)
\end{equation}
Hereafter the twisting element $\mathcal{F}$ is symbolically written in the
following form:
\begin{equation}
\ \mathcal{F}=\mathrm{f}^\alpha\otimes\mathrm{f}_\alpha \qquad \mbox{and}
\qquad \mathcal{F}^{-1}=\bar{\mathrm{f}}^\alpha\otimes\bar{\mathrm{f}}_\alpha
\end{equation}
Quantized Hopf algebra has nondeformed algebraic sector (commutators),
while coproducts and antipodes are subject to deformation:
\begin{equation}
\Delta^{\mathcal{F}} (\cdot)=\mathcal{F} \Delta (\cdot)\mathcal{F}^{-1}
,\quad S^{\mathcal{F}}(\cdot)=u\,S(\cdot)\,u^{-1}
\end{equation}
where $u=\mathrm{f}^\alpha S(\mathrm{f}_\alpha)$. Let us recall that
twisting two-tensor $\mathcal{F}$ is an invertible element in $U_{\mathcal{X}%
\mathds{M}}\otimes U_{\mathcal{X}\mathds{M}}$ which fulfills the 2-cocycle and normalization
conditions \cite{Drinfeld,Chiari}:
\begin{equation}  \label{coc}
\ \mathcal{F}_{12}(\Delta\otimes id)\mathcal{F}=\mathcal{F}%
_{23}(id\otimes\Delta)\mathcal{F}\ ,\quad \ (\epsilon\otimes id)\mathcal{F}%
=1=(id\otimes\epsilon)\mathcal{F}
\end{equation}
Its relation with a corresponding classical $r$ -matrix $\mathfrak{r}$
satisfying the classical Yang-Baxter equation is via a universal (quantum) $r$ -
matrix $\mathcal{R}$:
\begin{equation}  \label{qR}
\mathcal{R}=\mathcal{F}^{21}\mathcal{F}^{-1}=1+a\, \mathfrak{r}\, \ \
mod(a^2)
\end{equation}
where $a$ denotes the deformation parameter. As it is well -known from the general
framework of quantum deformations \cite{Drinfeld}, a twisted deformation
requires a topological extension of the enveloping algebra $U_{\mathfrak{g}}$
of some Lie algebra $\mathfrak{g}$ into an algebra of formal power series $%
U_{\mathfrak{g}}[[a]]$ in the formal parameter $a$ (see, e.g., \cite{Bonneau,Chiari}).
For the purpose of the present paper, we shall call $U_{\mathcal{X}\mathds{M}}[[a]]$
an algebra of formal (or generalized) differential operators. Accordingly,
the Hopf module algebra $\mathcal{A}_{\mathds{M}}$ has to be extended to $\mathcal{A}%
_{\mathds{M}}[[a]]$ as well. Particularly, deformed algebra $\mathcal{A}_{\mathds{M}}^{%
\mathcal{F}}$ can be represented by deformed $\star-$commutation relations
\begin{equation}\label{tcr}
\ [x^{\mu },x^{\nu }]_{\star }\equiv x^{\mu }\star x^{\nu }-x^{\nu
}\star x^{\mu }=\theta^{\mu\nu}(x)\equiv
\imath\theta^{\mu\nu}+\imath\theta^{\mu\nu}_\lambda x^\lambda+\dots
\end{equation}
replacing the nondeformed (commutative) one
\begin{equation}
\ [x^{\mu },x^{\nu }]=0
\end{equation}
where the coordinate functions $(x^\mu)$ play a role of generators for the
corresponding algebras:  deformed and nondeformed. The action of
differential operators on functions induced by derivations (vector fields)
remains the same in deformed and nondeformed cases. Moreover, the twisted star
product enables us to introduce two operator realizations of the algebra $%
\mathcal{A}_{\mathds{M}}^{\mathcal{F}}$ in terms of (formal) differential operators
on $\mathds{M}$. The so-called left-handed and right-handed realizations are
naturally defined by
\begin{equation}  \label{lhr-rhr1}
\hat{x}^{\mu }_L (f)= x^{\mu }\star f \qquad \mbox{and} \qquad \hat x^{\mu
}_R (f)= f\star x^{\mu }
\end{equation}
with $\hat{x}^{\mu }_L $, $\hat{x}^{\mu }_R$ $\in U_{\mathcal{X}\mathds{M}}[[a]]$
satisfying the operator commutation relations
\begin{equation}  \label{lhr-rhl2}
[\hat{x}^{\mu }_L, \hat{x}^{\nu }_L ]=\theta^{\mu\nu}(\hat{x})\qquad %
\mbox{and} \qquad [\hat{x}^{\mu }_R, \hat{x}^{\nu }_R ]=-\theta^{\mu\nu}(%
\hat{x})
\end{equation}
correspondingly. In other words, the above formulas describe embedding of $%
\mathcal{A}_{\mathds{M}}^{\mathcal{F}}$ into $U_{\mathcal{X}\mathds{M}}[[a]]$. These operator
realizations are particular cases of the so-called Weyl map (see, e.g., \cite{Weyl} and references therein for more details). They allow us to calculate the commutator:
\begin{equation}  \label{cc}
[x^\mu, f]_\star=\hat{x}^{\mu }_L (f)-\hat{x}^{\mu }_R(f)
\end{equation}

It has been argued in Ref. \cite{JSSW} (see also \cite{BMS}) that any Lie-algebraic star product (when generators satisfy the Lie algebra structure)
\begin{equation}  \label{c1}
\ [x^{\mu },x^{\nu }]_{\star }=\imath\theta_{\lambda }^{\mu \nu }x^{\lambda }
\end{equation}
can be obtained by twisting an element in the form
\begin{equation}  \label{c2}
\ \mathcal{F}= exp(\frac{\imath}{2}x^\lambda g_\lambda(\imath\frac{\partial}{%
\partial y},\imath\frac{\partial}{\partial z}))\ \mid_{y\shortrightarrow x;\
z\shortrightarrow x}
\end{equation}
The star product
\begin{equation}  \label{c3}
\ f(x)\star g(x)=exp(\frac{\imath}{2}x^\lambda g_\lambda(\imath\frac{\partial%
}{\partial y},\imath\frac{\partial}{\partial z}))f(y)g(z) \mid_{y%
\shortrightarrow x; z\shortrightarrow x }
\end{equation}
would imply
\begin{equation}  \label{c4}
[x^\mu, f]_\star=\imath\theta_{\lambda }^{\mu \nu }x^{\lambda }\partial_\nu f
\end{equation}
i.e., a vector field action on $f$. The last formula is particularly
important for obtaining a Seiberg-Witten map for noncommutative gauge
theories (see \cite{JSSW,BMS}). We are going to show using explicit
examples that this formula is not satisfied for an arbitrary twist in
the form of (\ref{c2}). However, we shall find an explicit twist for
$\kappa-$deformed Minkowski spacetime which belongs to the class
described by (\ref{c4}).

Before proceeding further, let us comment on some important
differences between the canonical and Lie-algebraic cases. In the
former, related to the Moyal product case
\begin{equation}  \label{rc1}
[x^\mu, x^\nu]_{\star, M}=\imath\theta^{\mu \nu }
\end{equation}
with a constant antisymmetric matrix $\theta^{\mu \nu }$; cf.
(\ref{tcr}), one finds that
\begin{equation}  \label{rc2}
[g, f]_{\star, M}=\partial_\nu\zeta^{\nu }(f,g)
\end{equation}
is a total derivative, e.g.,
\begin{equation}  \label{rc3}
[x^\mu, f]_{\star, M}=\partial_\nu\left(\imath\theta^{\mu \nu}f
\right)
\end{equation}
This further implies the following tracial property of the integral:
\begin{equation}  \label{rc4}
\int d^n\,x\, [g, f]_{\star, M}= 0
\end{equation}
which is rather crucial for  a variational derivation of Yang-Mills
field equations (see \cite{ADMSW}). In contrast to (\ref{rc3}),
Eq. (\ref{c4}) rewritten under the form
\begin{equation}  \label{rc5}
[x^\mu, f]_\star=  \partial_\nu\left(\imath\theta^{\mu \nu}_\lambda
x^\lambda f \right)- \imath\theta_{\nu }^{\mu \nu } f
\end{equation}
indicate  obstructions to the tracial property (\ref{rc4}) provided
that $\theta_{\nu }^{\mu \nu }\neq 0$. \footnote{For  the
$\kappa$ -deformation [see (\ref{kM}) and (\ref{c5}) below], one has
$\theta_{\nu }^{\mu \nu }=(n-1)a$.}

\section{$\protect\kappa-$Minkowski spacetime from twist: Hopf module
algebra point of view}

Our first task is to find explicit twists in order to achieve a twisted star-product
realization of the well-known $\kappa -$deformed Minkowski spacetime algebra
\cite{Z,MR}:
\begin{equation}  \label{kM}
\ [x^{0},x^{m}]_{\star }=\imath a x^{m}\equiv\frac{\imath}{\kappa }x^{m}
\end{equation}
for $m=1,...,n-1$ with remaining elements commuting. Here $a$ is the above-mentioned  formal parameter and $\kappa=\frac{1}{a}$ has the mass
dimension. Strictly speaking, formulas (\ref{kM}) mean that the
corresponding algebra of functions $\mathcal{A}_{\mathds{M}}[[a]]$ has been provided
with a twisted star product (\ref{tsp}) which leads to the commutation
relations (\ref{kM}). Of course, as we shall see later on, different twisted
star products may lead to the same commutation relations (\ref{kM}). In what
follows, we shall present explicit results for two one-parameter families of
twists providing quantum $\kappa-$Minkowski spacetime.

\subsection*{Jordanian family}

To this aim we shall consider a one-parameter family of two-dimensional
Borel subalgebras $\mathfrak{b}^{2}(r)=\{J_r,P_{r}\}\subset \mathfrak{igl}(n,%
\mathds{R})$:
\begin{equation}  \label{borel}
[J_{r},P_{r}]=P_{r }\ ,\quad
\ J_{r }=-x^{0}\partial _{0}+\frac{1}{r}x^{k}\partial _{k} , \quad
P_{r}=r\partial _{0}
\end{equation}%
with a numerical factor $r\neq 0$.
In terms introduced before the $\mathfrak{sl}(n)$ basis, the element $J_r$
has the form $J_{r}=\left(\frac{n-1}{r}-1\right)L-\frac{r+1}{r}L_0$. The
corresponding one-parameter family of Jordanian twists\footnote{%
The Borel subalgebra commutation relation leads to the validity of the
cocycle condition (\ref{coc}) which in turn guarantees associativity of the
corresponding star product.} can be expressed as (see \cite{Ogievetsky}--\cite%
{VNT1} for Jordanian twists):
\begin{equation}  \label{Jt}
\ \mathfrak{J}_{r }=\exp \left(J_{r}\otimes \sigma_r \right)
\end{equation}%
where $\sigma_r =\ln (1+\imath arP_0)$ with formal parameter $a=\frac{1}{%
\kappa }$.
Direct calculations show that, regardless of the value of $r$, twisted
commutation relations (\ref{tcr}) take the form of that for $\kappa-$Minkowski spacetime
(\ref{kM}).

The twists (\ref{Jt}) can be used to deform the entire $U_{\mathcal{X}\mathds{M}}$ Hopf
algebra. For generic $r\neq 0$, the smallest subalgebra containing
simultaneously the Borel subalgebra (\ref{borel}) and one of the orthogonal
subalgebras $\mathfrak{iso}(g; n)$ [e.g., Poincar\'{e} subalgebra $\mathfrak{iso}%
(n-1, 1)$] is $\mathfrak{igl}(n)$. However, there are three exceptions.%
\newline

(A) For $r=n-1$ in $n-$dimensional spacetime, the smallest subalgebra is $%
\mathfrak{isl}(n)$. \newline

(B) For $r=-1$ ($J_{-1}=-L$) in an arbitrary dimension, the smallest subalgebra
is Weyl-orthogonal algebra $\mathfrak{iwso}(g; n)$. It contains a central
extension of any orthogonal algebra $\mathfrak{so}(g; n)$. \footnote{%
The signature of the metric $g$ is irrelevant from an algebraic point of view.} In
this case, the commutation relation (\ref{isog1}-\ref{isog2}) should be
supplemented by
\begin{equation}  \label{iwso}
[M_{\mu\nu}, L]=0, \qquad\qquad [P_\mu, L]= P_\mu
\end{equation}
Of course, for physical applications we will choose the Weyl-Poincar\'{e}
algebra. This minimal one-generator extension of the Poincar\'{e} algebra $\mathfrak{iso}(n-1, 1)$ has
been used in Ref. \cite{Ballesteros} (cf. Appendix B).\newline

(C) $r=1$ in $n=2$ dimensions, $J_1=M_{10}$ is a boost generator for
nondiagonal metric $g_{00}=g_{11}=0$, $g_{01}=g_{10}=1$ with the Loerentzian signature. This corresponds
to the so-called light-cone deformation of the Poincar\'{e} algebra $\mathfrak{%
iso}(1, 1)$ \cite{LLM}.\newline

Since in the generic case we are dealing with $\mathfrak{igl}(n)$ Lie algebra, we
shall write deformed coproducts and antipodes in terms of its generators $%
\{L_{\nu }^{\mu },P_{\mu }\}$. The deformed coproducts read as follows:
\begin{equation}
\Delta _{r}\left( P_{0}\right) =1\otimes P_{0}+P_{0}\otimes e^{\sigma_r },
\qquad \Delta _{r}\left( P_{k}\right) =1\otimes P_{k}+P_{k}\otimes e^{-\frac{%
1}{r}\sigma_r }
\end{equation}
\begin{equation}
\Delta _{r}\left( L_{k}^{m}\right) =1\otimes L_{k}^{m}+L_{k}^{m}\otimes 1,
\qquad \Delta _{r}\left( L_{0}^{k}\right) =1\otimes
L_{0}^{k}+L_{0}^{k}\otimes e^{\frac{r+1}{r}\sigma_r }
\end{equation}
\begin{equation}
\ \Delta _{r}\left( L_{k}^{0}\right) =1\otimes L_{k}^{0}+L_{k}^{0}\otimes
e^{-\frac{r+1}{r}\sigma_r }-\imath arJ_r\otimes P_{k}e^{-\sigma_r }
\end{equation}
\begin{equation}
\ \Delta _{r}\left( L^0_0\right) =1\otimes L^0_0+L^0_0\otimes 1\ -\imath ar
J_r\otimes P_{0}e^{-\sigma_r }
\end{equation}
Here
\begin{equation*}
e^{\beta\sigma_r}=\left(1+ \imath arP_0\right)^\beta=\sum_{m=0}^\infty \frac{%
a^l}{m!}\beta^{\underline m}(\imath rP_0)^m
\end{equation*}
and $\beta^{\underline m}=\beta(\beta-1)\dots (\beta-m+1)$ denotes the
so-called falling factorial. The antipodes are:
\begin{equation}
S_r\left( P_{0}\right) =-P_{0}e^{-\sigma_r }, \qquad S_r\left( P_{k }\right)
=-P_{k }e^{\frac{1}{r}\sigma_r }
\end{equation}
\begin{equation}
S_r\left( L_{0}^{k}\right) =-L_{0}^{k}e^{-\frac{r+1}{r}\sigma_r}, \qquad
S_r\left( L_{k}^{0}\right) =-\left(L_{k}^{0}+\imath arJ_rP_k\right)e^{\frac{%
r+1}{r}\sigma_r }
\end{equation}
\begin{equation}
S_r\left( L_{0}^{0}\right) =-L_{0}^{0}-\imath arJ_rP_0, \qquad S_r\left(
L_{k}^{m}\right)=-L_{k}^{m}
\end{equation}

The Jordanian one-parameter family of twists (\ref{Jt}) generates, of course,
left- and right-hand representations, respectively, which give a realization of
(\ref{lhr-rhr1}) and (\ref{lhr-rhl2}) in the following form:\newline
I. Left-handed representations :
\begin{equation}  \label{Jtlh}
\ \hat{x}^{i}_{L,r}=x^{i}\left( 1+rA\right)^{-\frac{1}{r }} ,\qquad %
\mbox{and}\qquad \ \hat{x}^{0}_{L,r}=x^{0}(1+rA)
\end{equation}%
Hereafter one introduces for convenience a Hermitian operator $%
A=ia\partial_0\equiv\frac{i}{\kappa}\partial_0$ . \footnote{%
This notation will be particularly convenient and utilized in the subsequent
section.} \newline
II. Right-handed representations (Hermitian for $r=n-1$):
\begin{equation}  \label{Jtrh}
\ \hat{x}^{i}_{R,r}=x^{i} , \qquad \mbox{and}\qquad \hat{x}%
^{0}_{R,r}=x^{0}(1+r A)-\imath ax^{k}\partial _{k}
\end{equation}%
Particularly, using (\ref{cc}), we obtain
\begin{equation}
[x^i, f]_{\star,r}= x^{i}\left(\left( 1+rA\right)^{-\frac{1}{r }}-1\right)f
,\qquad [x^0, f]_{\star,r}= \imath a x^k\partial_k f
\end{equation}
which is different from (\ref{c1}). However for $r=-1$ one obtains the desired
commutator
\begin{equation}\label{c5}
[x^\mu, f]_\star=\imath\left( a^{\mu}x^\nu-a^{\nu}x^\mu\right)\partial_\nu f
\end{equation}
providing the $\kappa-$deformed Minkowski spacetime, i.e., $%
a^\mu=(a,0,\ldots,0)$.
\subsection*{Abelian family}

$\kappa -$Minkowski spacetime can be also implemented by the one-parameter
family of Abelian twists \cite{Bu, MSSKG} (with $s$ being a numerical
parameter):
\begin{equation}  \label{At}
\ \mathfrak{A}_{s}=\exp \left[ -\imath a\left(s\partial _{0}\otimes
x^{k}\partial _{k}-\left( 1-s\right) x^{k}\partial _{k}\otimes \partial
_{0}\right)\right]
\end{equation}
All are $U_{\mathfrak{igl}(n)}$ twists in a sense explained before. A
special case $s=\frac{1}{2}$ has been treated in Ref. \cite{Bu}. Thus one gets the following:
\newline
I. Left-handed representation (Hermitian for $s=0$) :
\begin{equation}  \label{Atlh}
\hat x^{i}_{L,s}=x^{i}e^{\left( s-1\right) A} , \qquad \hat x^{0}_{L,s} =
x^{0}+\imath asx^{k}\partial _{k}
\end{equation}%
II. Right-handed representation (Hermitian for $\ s=1$):
\begin{equation}  \label{Atrh}
\hat x^{i}_{R,s}=x^{i}e^{sA} , \qquad \hat x^{0}_{R,s}=x^{0}+\imath a\left(
s-1\right) x^{k}\partial _{k}
\end{equation}
This implies
\begin{equation}
[x^i, f]_{\star,s}= x^{i}\left(e^{(s-1)A}-e^{sA}\right)f ,\qquad [x^0,
f]_{\star,s}= \imath a x^k\partial_k f
\end{equation}
which is different from (\ref{c1}) for any value of the parameter $s$.
The deformed coproducts read as follows (cf. \cite{Bu}):
\begin{equation}
\Delta _{s}\left( P_{0}\right) =1\otimes P_{0}+P_{0}\otimes 1,\qquad \Delta
_{s}\left( P_{k}\right) =e^{\imath asP_0 }\otimes P_{k}+P_{k}\otimes
e^{-\imath a(1-s)P_0}
\end{equation}
\begin{equation}
\Delta _{s}\left( L_{k}^{m}\right) =1\otimes L_{k}^{m}+L_{k}^{m}\otimes 1,
\qquad \Delta _{s}\left( L_{0}^{k}\right) =e^{-\imath asP_0 }\otimes
L_{0}^{k}+L_{0}^{k}\otimes e^{\imath a(1-s)P_0}
\end{equation}
\begin{equation}
\ \ \Delta _{s}\left( L_{k}^{0}\right) = e^{\imath asP_0 }\otimes
L_{k}^{0}+L_{k}^{0}\otimes e^{-\imath a(1-s)P_0} -\imath as P_k\otimes
D+\imath a(1-s) D\otimes P_k
\end{equation}
\begin{equation}
\ \Delta _{s}\left( L^0_0\right) = 1\otimes L^0_0+L^0_0\otimes 1\ +\imath as
P_0\otimes D-\imath a(1-s) D\otimes P_0
\end{equation}

Here $D=x^k\partial_k$. The antipodes are
\begin{equation}
S_s\left( P_{0}\right) =-P_{0}, \qquad S_s\left( P_{k }\right) =-P_{k
}e^{-\imath aP_0}
\end{equation}
\begin{equation}
S_s\left( L_{k}^{m}\right) =-L_{k}^{m}, \qquad S_s\left( L_{0}^{k}\right)
=-L_{0}^{k} e^{\imath aP_0}
\end{equation}
\begin{equation}
S_s\left( L_{0}^{0}\right) =-L_{0}^{0}-\imath a(1-2s)DP_0
\end{equation}
\begin{equation}
S_s\left( L_{k}^{0}\right) =-e^{-\imath asP_0}L_{k}^{0}e^{\imath
a(1-s)P_0}-\imath a\left[sP_kDe^{-\imath asP_0}-(1-s)DP_k e^{\imath
a(1-s)P_0}\right]
\end{equation}
The above relations are particularly simple for $s=0, 1, {\frac{1}{2}}$.

Before closing this section, let us point out that all twists considered here
correspond to the same classical $r$ -matrix (Poisson bi-vector on $\mathds{M}
$):
\begin{equation}
\mathfrak{r}=(x^k\partial_k)\wedge \partial_0\in {\mathcal X}\mathds{M}\wedge {\mathcal X}\mathds{M}
\end{equation}

\section{Lie algebra of $\protect\kappa-$deformed Lorentzian spacetime.}

The noncommutative $\kappa-$Minkowski space can be realized in quantized
relativistic phase space \cite{LRZ} or in a Schr\"{o}dinger representation ($%
x^\mu, p_\mu=-\imath\partial_\mu$) in terms of generalized differential
operators.

In Refs. \cite{MSSKG,MS,MS2}, Meljanac et al., following an earlier development \cite%
{DJMTWW,Dimitrijevic}, have found an interesting realization of the
noncommutative coordinates $\hat{x}^{\mu }$ in terms of generalized
differential operators. Their approach assumes the following ansatz:
\begin{equation}  \label{a1}
\ \hat{x}^{i}=x^{i}\phi (A) , \qquad\qquad \ \hat{x}^{0}=x^{0}\psi
(A)+\imath ax^{k}\partial _{k}\gamma (A)
\end{equation}%
for noncommutative coordinates $\hat{x}^{\mu }\in U_{\mathcal{X}\mathds{M}}[[a]]$,
where as before $A=ia\partial _{0}$. Functions $\phi $, $\psi $, and $\gamma $
are taken to be real analytic; however generalization to complex analytic is
straightforward and will not be discussed here.
These functions obey initial conditions $\phi (0)=1$ and $\psi
(0)=1$, and $\gamma (0)$ has to be finite in order to ensure a proper classical limit
at $a\mapsto 0$. The operators $\hat{x}^{i}$ are
automatically Hermitian while Hermiticity of $\hat{x}^{0}$ requires the
additional assumption that
\begin{equation}  \label{herm}
\psi ^{\,\prime}+(n-1)\gamma=0
\end{equation}%
where $\psi ^{\,\prime }\equiv\frac{d\psi }{d A}$ and $n$ denotes spacetime
dimension. As we will see later on, this condition will be satisfied only in a
few exceptional cases. Apparently, the ansatz (\ref{a1}) could be guessed from
our twisted realizations (\ref{Jtlh}), (\ref{Jtrh}), (\ref{Atlh}), (\ref{Atrh}).

Now, the $\kappa-$Minkowski commutation relations [cf. (\ref{kM})]
\begin{equation}  \label{kM2}
\ [\hat x^{0},\hat x^{k}]=\imath a\tau \hat x^{k}
\end{equation}
are equivalent to the property that functions $\phi ,\psi $, and $\gamma $ do
satisfy the ODE \cite{MS}:
\begin{equation}  \label{kMode}
(\ln\phi) ^{\prime }\psi =\gamma -\tau \
\end{equation}
hereafter $\tau =\pm 1$ for convenience [cf. (\ref{lhr-rhl2})]
\footnote{The sign convention $\tau=\pm 1$ is due to the difference between left- and right-handed realizations; see (\ref{lhr-rhl2}). In fact, it can be eliminated by
rescaling $\hat x^0\mapsto \tau\,\hat x^0$.}.

Throughout this paper we shall be interested in solutions of (\ref{kMode})
for $\gamma=constant$. In this case, for any $\psi$ one easily gets
\begin{equation}  \label{s-phi}
\phi=\Psi ^{\gamma-\tau} \qquad \mbox{where}\quad \Psi=\exp\left(\int_0^A%
\frac{dA^{\prime }}{\psi(A^{\prime })}\right)
\end{equation}
Specializing further to the linear case $\psi=1+rA$, one finds
\begin{equation}  \label{om-1}
\Psi =e^A \qquad \mbox{for}\quad r=0
\end{equation}
and
\begin{equation}  \label{om-2}
\Psi = (1+rA)^{\frac{1}{r}} \qquad \mbox{for}\quad r\neq 0
\end{equation}
The Hermiticity restriction (\ref{herm}) is  satisfied only for $\psi=1$ and $\gamma=0$
and in $n$ -dimensions for $\psi=1\pm (n-1)A, \gamma=\mp 1$ subcases. All of our
twisted products realizations turn out to be special cases of the above more
general formulas (\ref{s-phi})-(\ref{om-2}).

I.) One finds that the first case (\ref{om-1}), i.e., $\psi=1$, corresponds
to the Abelian twists with $\gamma=s, \tau=+1$ for left-handed realizations
[cf. (\ref{Atlh})] and $\gamma=s-1, \tau=-1$ for right-handed ones. Two
subcases $s=0, \tau=1$ and $s=1, \tau=-1$ give rise to Hermitian
representations.

II.) The case $r\neq 0$ (\ref{om-2}) is related to the Jordanian family under
rather restricted values of $\gamma$: for left-handed representation $%
\gamma=0, \tau=1$, while $\gamma=\tau=-1$ for right-handed ones [cf. (\ref%
{Jtrh})]. We do not know twist realizations for generic $\gamma$ and $r\neq 0
$. Among Jordanian twists only the case $\gamma=\tau=-1$ with $r=n-1$ is
Hermitian in a spacetime of dimension $n$ and can be reduced to the subgroup
$\mathfrak{isl}(n)$.

Now, following the general method developed in Ref. \cite{MS} (see also \cite{Dimitrijevic}), one
can try to covariantly incorporate the $\kappa-$Minkowski algebra (\ref{kM2}%
) into the extension of undeformed orthogonal algebra (\ref{isog1}) by
assuming
\begin{equation}  \label{L1}
\ [M_{\mu \nu }, M_{\rho \lambda }]=\eta _{\nu \rho } M_{\mu \lambda }+
\eta_{\mu \lambda }M_{\nu \rho }-\eta _{\nu \lambda }M_{\mu \rho }-\eta
_{\mu\rho }M_{\nu\lambda}
\end{equation}%
\begin{equation}  \label{L2}
\ \left[ M_{\mu\nu}, \hat{x}_\lambda\right] = \eta_{\nu\lambda}\hat{x}%
_{\mu}-\eta_{\mu\lambda}\hat{x}_\mu -\imath a_\mu M_{\nu\lambda}+\imath
a_\nu M_{\mu\lambda}
\end{equation}%
where $\hat{x}_\lambda=\eta_{\lambda\nu}\hat{x}^\nu$ and $%
a_\mu=\eta_{\mu\lambda}a^\lambda: (a^\nu)=(\tau\,a, 0,\ldots,0)$ . The main
point is the ansatz
\begin{equation}  \label{a2}
\ M_{i0}=x_{i}\partial _{0}F_{1}-x_{0}\partial _{i}F_{2}+\imath ax_{i}\Delta
F_{3}+\imath ax^{k}\partial _{k}\partial _{i}F_{4}
\end{equation}%
\label{a3} where $F_{p}\equiv F_{p}(A), p=1,2,3,4$ are (real) analytic functions to
be determined from (\ref{L1}-\ref{L2}). In turn, generators
\begin{equation}  \label{a4}
\ M_{ij}=x_{i}\partial _{j}-x_{j}\partial _{i}  \notag
\end{equation}
remain undeformed, i.e., 
in the Schwinger realization: $M_{ik}^*=-M_{ik}$. Formula (\ref{a2})
describes the deformed realization of the "boost" generators $M_{i0}$ together
with the initial conditions $F_1(0)=F_2(0)=1$. A difference between our and
the original approach \cite{MS} is that we do not apriori assume the Euclidean signature
for the metric $\eta_{\mu\nu}$.
One should notice that the Lorentzian signature $\eta_{\mu\nu}= (-1, 1,
\ldots, 1)$ is more natural and expected in this context: the commutation relations (\ref%
{kM2}) distinguish one of the variables $x^0$. In contrast, the Euclidean
signature puts all variables on equal footing.

In order to keep under control the difference between the Euclidean and Lorentzian cases, we
shall temporarily introduce a coefficient $\epsilon=\eta_{00}=\pm 1$.
(Notice that $x_0=\epsilon x^0$ and $[\partial_0, x_0]=\epsilon$.)

Inserting ansatz (\ref{a2}) into the algebra (\ref{L1}), we obtain
the following equations:
\begin{equation}  \label{F1}
\ F_{1}F_{2}+AF_{1}^{\prime }F_{2}+\epsilon AF_{1}F_{4}-2\epsilon
AF_{1}F_{3}=1
\end{equation}%
\begin{equation}  \label{F2}
\ 2F_{3}^{2}-\epsilon F_{3}^{\prime }F_{2}+F_{3}F_{4}=0
\end{equation}%
where $\epsilon =1$ (Euclidean case) or $\epsilon=-1$ (Lorentzian case).
Substituting (\ref{a1}) and (\ref{a2}) into
\begin{equation}
\ [M_{i0},\hat{x}_{0}]=\epsilon \hat{x}_{i}+\imath a\epsilon \tau M_{i0}
\end{equation}%
\begin{equation}
\ [M_{i0},\hat{x}_{j}]=-\delta _{ij}\hat{x}_{0}+\imath a\epsilon \tau M_{ij}
\end{equation}%
one obtains some overdetermined system of ODE; see Appendix A for
details. Its solutions can be recast into the form (here $%
\phi=\Psi^{\gamma-\tau}$) :
\begin{equation}  \label{F3}
\ F_{2}=\frac{\psi }{\phi };\ \ \ F_{3}= \frac{\epsilon \tau}{2\phi };\ \ \
F_{4}=- \frac{\epsilon\gamma }{\phi }
\end{equation}%
and
\begin{equation}  \label{F4}
\ F_{1}\psi +AF_{1}^{\prime }\psi -AF_{1}\left( \gamma +\tau \right) -\phi =0
\end{equation}
The last equation is consistent with (\ref{kMode}), (\ref{F1}-\ref{F3}) for
both values $\epsilon=\pm 1$. 
 Its solution
\begin{equation}
\ F_{1}(A)=\Psi^\gamma\frac{\Psi-\Psi^{-1}}{2A}\equiv\Psi^\gamma\frac{%
\Psi^\tau-\Psi^{-\tau}}{2A\tau}
\end{equation}%
together with (\ref{F3}) determines generators $M_{0i}$ completely. It is
worth noticing that the Hermticity of $\hat x^0$ automatically implies reality
for the boost generators: $M_{0i}^*=-M_{0i}$. Since the Euclidean case has
been already studied in Refs. \cite{MS,MS2}, further on, until the end of the present
paper,we shall use only Lorentzian signature $\epsilon=-1$.\newline

Analogously, following Ref. \cite{MS} we have also obtained (with some sign
corrections) a generalized d'Alambert operator $\tilde\square=\triangle
H_{1} -\partial _{0}^{2}H_{2}$:
\begin{equation}  \label{L3}
[M_{\mu\nu},\,\tilde\square]=0
\end{equation}
under the form [$H_1(0)=H_2(0)=1$]
\begin{equation}\label{L33}
H_1(A)=\Psi^{\tau-2\gamma}, \qquad H_2(A)=\left(\Psi+\Psi^{-1}-2\right)A^{-2}
\end{equation}
where $\triangle=\partial_k\partial^k$ above denotes a $n-1$ dimensional,
spacelike, Laplace operator.
We are now in a position to define Dirac derivatives $D_\mu$ as
\begin{equation}  \label{L4}
[\tilde\square, \hat{x}_\mu]=2D_\mu \ ,\qquad [\tilde\square, D_\mu]=0
\end{equation}
This implies
\begin{equation}
D_{0}=\partial _{0}G_{2} +\imath a\Delta G_{3} , \qquad D_{i}=\partial
_{i}G_{1}
\end{equation}
where (see Appendix A)
\begin{equation}  \label{G1}
G_1=H_1\phi=\Psi^{-\gamma} , \quad G_2=G_1F_1=\frac{\Psi-\Psi^{-1}}{2A} ,
\quad G_3=-\frac{\tau}{2}H_1=-\frac{\tau}{2}\Psi^{\tau-2\gamma}
\end{equation}
Let us remark that only formulas for $G_2$ and $H_2$ are universal in the sense
that they do not depend on the parameters $\gamma$ and $\tau$.

Direct calculations performed on solutions with constant $\gamma$ give rise
to
\begin{equation}  \label{L5}
D_\mu D^\mu =\tilde\square\left(1+\frac{a^2}{4}\tilde\square\right), \quad
\sqrt{1+a^2D_\mu D^\mu}=1+\frac{a^2}{2}\tilde\square
\end{equation}
and ($\tau=\pm 1$)
\begin{equation*}
\Psi^{-\tau}=-\imath a\tau D_0+\sqrt{1+a^2D_\mu D^\mu} .
\end{equation*}%
These relations allow us to enlarge the Lorentz - $\kappa-$Minkowski algebra
(\ref{kM2}), (\ref{L1}), (\ref{L2}) by the following commutation relations:
\begin{equation}  \label{L6}
[D_\mu, D_\nu]=0 , \qquad [M_{\mu\nu},
D_\lambda]=\eta_{\nu\lambda}D_\mu-\eta_{\mu\lambda}D_\nu
\end{equation}
\begin{equation}  \label{L7}
[D_k, \hat x_0]=0 , \qquad [D_k, \hat x_j]=\delta_{jk}\left(-\imath a\tau
D_0+\sqrt{1+a^2D_\mu D^\mu}\right)
\end{equation}
\begin{equation}  \label{L8}
[D_0, \hat x_j]=-\imath a\tau D_j , \qquad [D_0, \hat x_0]=-\sqrt{1+a^2D_\mu
D^\mu}
\end{equation}
in order to include the Dirac derivatives. In this way one has obtained a
new (non Lie-algebraic) quantum extension of the Lorentz algebra which
contains $\frac{1}{2}n(n+3)$ generators $(M_{\mu\nu}, \hat x_\mu, D_\mu )$
in $n$ -dimensional spacetime $\mathds{M}$
\footnote{It is a relativistic version of the Euclidean algebra obtained already in \cite{MS}.}.
Its algebraic structure is
completely described by the commutation relations (\ref{kM2}), (\ref{L1}), (\ref{L2}), (\ref%
{L6}) - (\ref{L7}) and does not depend on a particular differential
operator realization which had been used for its construction.
Particularly, it does not depend on a twisting tensor itself. It
contains undeformed Poincar\'{e} algebra $\mathfrak{iso}(n-1, 1)$:
$(M_{\mu\nu},D_\mu )$; cf. (\ref{L6}). Another Lie-algebraic part
$(M_{\mu\nu}, \hat x_\mu)$ splits into Lorentzian subalgebra
$\mathfrak{so}(n-1,1)$ generated by $(M_{\mu\nu})$ combined with
the quantum $\kappa-$Minkowski space $(\hat x^{\mu})$: (\ref{kM2}), (\ref{L1}%
), (\ref{L2}). It contains one free (formal) parameter $a$ and sign convention $%
\tau=\pm 1$. From the algebraic point of view this dependence can be removed
by rescaling $\hat x_0\mapsto \frac{\tau}{a}\hat x_0$ or equivalently by
setting $\tau a=1$. It makes this algebra isomorphic by substituting
\begin{equation}  \label{dS}
\hat x_0=\imath a\,\tau\,M_{0n}\ ,\qquad\qquad \hat x_i\sim M_{0i}-M_{in}
\end{equation}
to the (nondeformed) simple $\mathfrak{so}(n, 1)$ Lie algebra for a "frozen" value of the parameter $a$.
This observation can be helpful  for studying representations for such systems.

From the physical point of view, however, a dimensionful constant $\kappa={%
\frac{1}{a}}$ can be related with some fundamental constant of nature,
similarly like in DSR \footnote{%
See \cite{JKG1} for interrelations between DSR and de Sitter group $SO(4,1)$.%
}~. Assuming for a moment that twist has a physical meaning, we will see how
its parameters enter the so-called dispersion relations obtained from plane
wave solutions of the corresponding Klein-Gordon equations with a deformed
operator $\tilde\square$. Before doing that, let us emphasize that the presented
formalism has a well-defined classical limit $a\mapsto 0$ which reconstructs
standard Minkowski spacetime together with the Poincar\'{e} group acting on it and
Heisenberg-type relations between position and momenta operators.
Particularly, $\tilde\square$ becomes a standard d'Alambert operator $\square
$.

\subsection*{Dispersion relations}

To this aim let us consider as a toy model involving d'Alambert operator
\begin{equation*}
\tilde\square=\triangle H_1(A)-\partial_0^2 H_2(A)
\end{equation*}
in specific realizations (\ref{L33}). One is looking for a plane wave solution of the
deformed Klein-Gordon equation
\begin{equation}  \label{dAeq}
\left(\tilde\square - m_0^2\right)\omega_k=0
\end{equation}
where $\omega_k=\exp{(\imath\, k_\mu x^\mu)}$ represents the plane wave with the
covariant wave vector $k=(k_\mu)$; $m_0$ denotes a mass. Straightforward and
simple calculations give rise to the general form of deformed "dispersion
relation"
\begin{equation}
m_0^2= k_0^2 H_2\left(-\frac{k_0}{\kappa}\right)- \mathbf{k}^2 H_1\left(-%
\frac{k_0}{\kappa}\right)
\end{equation}
where $\mathbf{k}^2=k^ik_i$. Further specialization to the case when $\psi=1$
which corresponds to the Abelian twists give rise to
\begin{equation}
m_0^2= \left[2\kappa\sinh\left(\frac{k_0}{2\kappa}\right)\right]^2- \mathbf{k%
}^2 \exp\left[(2s-1)\frac{k_0}{\kappa}\right]
\end{equation}
For the case $s=1$, the above expression is in agreement with the formula
known from DSR and which had been originated in a $\kappa-$Poincar\'{e} algebra
deformed Casimir operator. However, for the Hermitian case $s=0$, the
dispersion formula is different. For Jordanian twists ($\psi=1+rA,\
\tau-2\gamma=1$), one gets instead
\begin{equation}
m_0^2= \kappa^2\left[ \left(1-r\frac{k_0}{\kappa}\right)^{\frac{1}{r}%
}+\left(1-r\frac{k_0}{\kappa}\right)^ {-\frac{1}{r}}-2\right]-\mathbf{k}^2
\left(1-r\frac{k_0}{\kappa}\right)^{\frac{1}{r}}
\end{equation}
Particularly, the Hermitian case in $n=4$ dimensions requires $r=3$ and
gives
\begin{equation}
m_0^2= k_0^2\left[1 +3\frac{k_0}{\kappa}+\frac{25}{3}\left(\frac{k_0}{\kappa}%
\right)^2 +\ldots\right]-\mathbf{k}^2 \left[1-\frac{k_0}{\kappa}-\left(\frac{%
k_0}{\kappa}\right)^2-\ldots\right]
\end{equation}
In contrast, for our special $r=-1$ case, the dispersion formula takes a very
simple linear "mass renormalization" form
\begin{equation}
m_0^2 \left( 1+\frac{k_0}{\kappa}\right)= k_0^2 - \mathbf{k}^2
\end{equation}
Its mathematical structure is similar to that in Ref. \cite{Smolin}.
Particularly, for  (free) massless fields, e.g., in electrodynamics, wave equations
remain unchanged.

\section{Conclusion}

We have obtained $\kappa$-deformed Minkowski spacetime by twisting $%
\mathfrak{igl}(n,\mathds{R})$ with a one-parameter family of Jordanian and
Abelian twists.
In both cases, we have introduced deformed coproducts, antipodes, and
(generalized) differential operator realizations of noncommutative
coordinates $\hat{x}^\mu$. Hermitian representations are found for both
families of twists.

Deformed generators $M_{\mu\nu}, \hat x_\mu$, d'Alambert operator $%
\tilde\square$, and Dirac derivatives $D_\mu$ are found, following the method
developed in Refs. \cite{MS,MS2,Dimitrijevic}, to form a new (quantum) algebra of non-Lie type which
contains undeformed Poincar\'{e} algebra $\mathfrak{iso}(n-1, 1)$. Its Lie-algebraic part $(M_{\mu\nu}, \hat x_\mu)$ is shown to be isomorphic to the
classical simple Lie algebra $\mathfrak{so}(n,1)$.

As a physical application, we have calculated dispersion relations ensuing
from the plane wave solutions of the generalized Klein-Gordon equation. One
recovers a standard DSR dispersion formula as well as a new one. Although
such twist dependence of dispersion relations seems to be unsatisfactory
from a physical point of view, in principle one does not know what to expect
at the Planck scale. On the other hand, a similar situation is encountered in
quantum mechanics: a spectrum of the Schr\"{o}dinger operator depends on a
concrete realization (e.g., potential).

The special case of the Jordanian family gives the minimal symmetry algebra:
Weyl-Poincar\'{e} algebra which is studied in more details in Appendix B.
The dispersion formula related to this minimal case has a new and interesting
form of linear mass deformation. In addition, the commutation relations (\ref{tcr})
take a form (\ref{c5}) which is necessary in NC Yang-Mills theory.

We summarize the paper with a claim that the role of twist has to be
reconsidered: although infinitely many twists accomplish the same $\kappa-$%
Minkowski spacetime together with the same $\mathfrak{so}(n, 1)$ algebra and
its nonlinear extension, the physical properties of a concrete model are
twist-dependent. One possible answer may be done by relating twist with some
interaction. In order to solve this problem, some further studies and
discussions are needed.

\subsection*{Acknowledgements}

This paper has been supported by MNiSW Grant No. NN202 318534 and the
Bogliubov-Infeld Program. The authors acknowledge financial support from the Quantum Gravity network of the European Science Foundation, A.B. in "QG2 2008 Quantum Geometry and Quantum Gravity
Conference", Nottingham, UK~, 30th June - 4th July 2008 and A.P. in "New Paths Towards
Quantum Gravity", Holbaek, Denmark, 12th-16th May, 2008. Both of us acknowledge helpful discussions with P. Aschieri and M. Dimitrijevi\'{c}. Special thanks are due to J.
Lukierski and V.N. Tolstoy for valuable comments and reading the manuscript.

\section*{Appendix A}


Following the same methods as in Ref. \cite{MS} and taking into account the
Lorentzian signature for the spacetime metric, one obtains a partially modified
set of ODEs for functions $G_i$ determining generators:
\begin{equation*}
D_{i}=\partial _{i}G_{1}\left( A\right) ;\ D_{0}=\partial _{0}G_{2}\left(
A\right) +\imath a\Delta G_{3}\left( A\right) ;
\end{equation*}
and for functions $H_{i}$ introduced in the ansatz for the d'Alambert operator:
\begin{equation*}
\tilde{\square} =\triangle H_{1}\left( A\right) -\partial
_{0}^{2}H_{2}\left( A\right)
\end{equation*}
This set of equations reads
\begin{equation}
\ G_{1}F_{1}=G_{2}\ ;\quad \ F_{3}G_{1}=G_{3}
\end{equation}%
\begin{equation}
\ G_{1}^{\prime }F_{2}+ F_{4}G_{1}=0
\end{equation}%
\begin{equation}
\ AG_{2}^{\prime }F_{2}+F_{2}G_{2}+2 AF_{1}G_{3}-G_{1}=0
\end{equation}%
\begin{equation}
\ G_{3}^{\prime }F_{2}+2 G_{3}(F_{3}+F_{4})=0
\end{equation}
\begin{equation}
\ AH_{2}^{\prime }F_{2}+2F_{2}H_{2}-2F_{1}H_{1}=0
\end{equation}%
\begin{equation}
\ H_{1}^{\prime }F_{2}+2(F_{3}+F_{4})H_{1}=0
\end{equation}

\section*{Appendix B}

\subsection*{Minimal case: Weyl- Poincar\'{e} algebra}

The minimal case $r=-1$ in physical $n=4$ dimensions deserves special
attention. It is related to research in Ref. \cite{Ballesteros}, where, however, the $\kappa-$Minkowski spacetime
has been obtained with different techniques (see the introduction). Below we shall present
coproducts and antipodes for all 11
generators in "physical" basis $(M_{k}, N_k, L, P_\mu)$ of the Poincar\'{e}-Weyl
algebra containing the Lorentz subalgebra of rotation $M_k=-\frac{\imath}{2}%
\epsilon_{klm}M_{lm}$ and boost $N_k=\imath M_{k0}$ generators:
\begin{eqnarray}  \label{sP1}
[M_i,\,M_j ]\ =\ \imath\,\epsilon_{ijk}\,M_k ~,\qquad [M_i,\,N_j]\ =\
\imath\,\epsilon_{ijk}\,N_k ~,\qquad [N_i,\,N_j]\ =\ -\imath\,
\epsilon_{ijk}\,M_k~
\end{eqnarray}
Abelian four-momenta $P_\mu=-\imath\partial_\mu$ ($\mu=0,\dots,3\ , k=1,2,3$)%
\begin{eqnarray}  \label{P3}
[M_j,\,P_k]\!\!&=\!\!&\imath\,\epsilon_{jkl}\,P_l~,\qquad [M_j,\,P_0]\;=\;0~,
\\[5pt]
[N_j,\,P_k]\!\!&=\!\!&-\imath\,\delta_{jk}\,P_0~,\quad\;\;[N_j,\,P_0]\;=\;-%
\imath\, P_j^{}~.
\end{eqnarray}
and dilatation generator $L=x^\mu\partial_\mu$ as before:
\begin{equation}
[M_k,\,L]=[N_k,\,L]=0 \,\qquad [P_\mu,\,L]= P_\mu
\end{equation}
The deformed coproducts are
\begin{equation}
\tilde{\Delta}\left( P_{\mu}\right) =1\otimes P_{\mu}+P_{\mu}\otimes e^{%
\tilde{\sigma} }, \qquad \tilde{\Delta}\left( M_{k}\right) =1\otimes
M_{k}+M_{k}\otimes 1
\end{equation}
\begin{equation}
\tilde{\Delta}\left( N_{k}\right) =1\otimes N_{k}+N_{k}\otimes 1+\frac{1}{%
\kappa} L\otimes P_{k}e^{-\tilde{\sigma} }
\end{equation}
\begin{equation}
\tilde{\Delta}\left( L\right) =1\otimes L+L\otimes 1\ +\frac{1}{\kappa}
L\otimes P_{0}e^{-\tilde{\sigma} }
\end{equation}
Here $e^{-\tilde{\sigma}}=\left(1+\frac{1}{\kappa}P_0\right)^{-1}$ and $%
L=-J_{-1}$ . The antipodes are
\begin{equation}
\tilde{S}\left( P_{\mu}\right) =-P_{\mu}e^{-\tilde{\sigma} }, \qquad \tilde{S%
}\left( M_{k}\right) =-M_{k}
\end{equation}
\begin{equation}
\tilde{S}\left( N_{k}\right) =-N_{k}+\frac{1}{\kappa}LP_k\ , \quad \tilde{S}%
\left( L\right) =-L +\frac{1}{\kappa}LP_0
\end{equation}

The corresponding natural left- and right-handed representations of
noncommutative coordinates $\hat x_\mu$ and deformed Lorentz generators $%
M_{\mu\nu}$, Dirac derivatives $D_\mu$ and d'Alembert operator $\tilde\square
$ are determined by functions $\psi, \phi, F_i, G_i$, and $H_i$. \newline
In this case one gets (formulas below are spacetime dimension-independent) the following:%
\newline
I. For left-handed representation ($\tau =1, \gamma=0, r=-1$)
\begin{equation}
\psi =\phi= 1-A\ ,\ \ \ F_{1}(A)=\frac{2-A}{2\left( 1-A\right) }
\end{equation}
\begin{equation}
\ F_{2}=1 \ ,\ \ \ F_{3}=-\frac{1}{2}\psi^{-1}\ ,\ \ \ F_{4}=0
\end{equation}
\begin{equation}
G_{1}=1,\quad G_{2}=\frac{ A-2}{2(A-1)} ,\quad G_{3}=-\frac{1}{2}\psi^{-1}
\end{equation}
\begin{equation}
H_{1}=H_2=\psi^{-1}
\end{equation}
II. For right-handed representation ($\tau =\gamma=r=-1$)
\begin{equation}
\psi= 1-A\ ,\ \ \ \phi=1\ ,\ \ \ F_{1}(A)=1-\frac{1}{2}A
\end{equation}
\begin{equation}
\ F_{2}=1-A \ ,\ \ \ F_{3}=\frac{1}{2}\ ,\ \ \ F_{4}=-1
\end{equation}
\begin{equation}
G_{2}=\frac{ A-2}{2(A-1)} ,\quad G_{3}=\frac{1}{2}\psi^{-1},\quad G_1=H_{1}=H_2=\psi^{-1}
\end{equation}

\end{document}